\documentclass[12pt]{article}
\usepackage{graphicx,amsmath,amssymb}
\newcommand{\cinst}[2]{$^{\mathrm{#1)}}$~#2\par}
\newcommand{\crefi}[1]{$^{\mathrm{#1)}}$}
\newcommand{\HRule}{\rule{0.4\linewidth}{0.3mm}}

\newcommand{\pt}{$p_{\rm T}$}
\newcommand{\jpsi}{J/$\psi$}
\newcommand{\psip}{$\psi^\prime$}
\newcommand{\ccbar}{$c \bar{c}$}
\newcommand{\upsAll}{$\Upsilon$}
\newcommand{\upsOneS}{$\Upsilon(1S)$}
\newcommand{\upsTwoS}{$\Upsilon(2S)$}
\newcommand{\upsThreeS}{$\Upsilon(3S)$}

\begin{document}

\begingroup
\thispagestyle{empty} \baselineskip=14pt
\parskip 0pt plus 5pt

\begin{center}
{\large EUROPEAN LABORATORY FOR PARTICLE PHYSICS}
\end{center}

\bigskip
\begin{flushright}
CERN--PH--EP\,/\,2010--xxx\\
May 17, 2010
\end{flushright}

\bigskip
\begin{center}
{\Large\bf \boldmath 
A new approach to\\[3mm]
quarkonium polarization studies}

\bigskip\bigskip

Pietro Faccioli\crefi{1},
Carlos Louren\c{c}o\crefi{2},
and Jo\~ao Seixas\crefi{1,3}

\bigskip\bigskip\bigskip
\textbf{Abstract}

\end{center}

\begingroup
\leftskip=0.4cm \rightskip=0.4cm
\parindent=0.pt

  Significant progress in understanding quarkonium production requires
  improved polarization measurements, fully considering the intrinsic
  multidimensionality of the problem.  We propose a frame-invariant
  formalism which minimizes the dependence of the measured result on
  the experimental acceptance, facilitates the comparison with
  theoretical calculations, and provides a much needed control over
  systematic effects due to detector limitations and analysis biases.
  This formalism is a direct and generic consequence of the rotational
  invariance of the dilepton decay distribution and is independent of
  any assumptions specific to particular models of quarkonium
  production.

\endgroup


\vfill
\begin{flushleft}
\HRule\\

\cinst{1} {Laborat\'orio de Instrumenta\c{c}\~ao e F\'{\i}sica Experimental de
  Part\'{\i}culas (LIP),\\ ~~~Lisbon, Portugal} 
\cinst{2} {CERN, Geneva, Switzerland}
\cinst{3} {Physics Department, Instituto Superior T\'ecnico (IST),
  Lisbon, Portugal}
\end{flushleft}
\endgroup

\newpage


Quarkonium polarization measurements should provide key information
for the understanding of quantum chromodynamics
(QCD)~\cite{bib:YellowRep-QWG}, with the competing mechanisms
dominating in the different theoretical approaches to quarkonium
production leading to very different predictions.  In particular,
non-relativistic QCD calculations~\cite{bib:NRQCD}, dominated by
colour-octet components, predict that, at Tevatron energies and
asymptotically high \pt, the directly produced \jpsi\ and \psip\
mesons should be almost fully \emph{transversely} polarized (angular
momentum component $J_z = \pm 1$) with respect to their own momentum
direction (helicity frame), while NLO calculations of colour-singlet
quarkonium production~\cite{bib:lansberg-HP08} indicate a strong
\emph{longitudinal} ($J_z = 0$) polarization component.

The present experimental knowledge on quarkonium polarization is also
contradictory and puzzling.  The slightly longitudinal 
prompt-\jpsi\ polarization measured by CDF~\cite{bib:CDFpol2} along
the helicity axis (\jpsi\ momentum direction in the center-of-mass of
the collision system) 
is in clear disagreement with the expectations of all existing models.
The fact that these recent measurements disagree with the results
previously published by the same experiment~\cite{bib:CDFpol1} adds
further confusion to the picture.  
Equally disturbing is the lack of continuity between fixed-target and
collider results~\cite{bib:pol}.
Bottomonium polarization should be easier to interpret theoretically,
given that the heavier bottom quark mass should satisfy the
non-relativistic approximation much better than the \ccbar\ case.
However, the \upsOneS\ Tevatron measurements~\cite{bib:upsCDFD0},
available in the helicity frame and extending to \pt\ values around
15~GeV$/c$, are contradictory: CDF indicates unpolarized production;
D0 indicates longitudinal polarization.  The discrepancy between the
two results cannot be justified by their different rapidity windows.
At lower energy and \pt, E866~\cite{bib:e866_Ups} has shown yet a
different polarization pattern: the \upsTwoS\ and \upsThreeS\ states
have \emph{maximal} transverse polarization with respect to the
direction of motion of the colliding hadrons (Collins-Soper
frame~\cite{bib:coll_sop}), with no significant dependence on
transverse or longitudinal momentum, while the \upsOneS\ is only
weakly polarized, indicating a dominant role of feed-down
contributions.

To clarify this intricate situation, improved measurements are needed.
So far, most experiments have presented results based on a fraction of
the physical information derivable from the data: only one
polarization frame is used and only the polar projection of the decay
angular distribution is studied.  These incomplete results prevent
model-independent physical conclusions.
Moreover, such partial descriptions of the observed physical processes
reduce the chances of detecting possible biases induced by
insufficiently controlled systematic effects.
The forthcoming LHC measurements, in particular, would benefit from an
improved formalism.
In Ref.~\cite{bib:pol} we emphasized the importance of approaching the
polarization measurement as a multidimensional problem, determining
the full angular distribution in more than one frame. In this letter
we show the existence of a frame-independent relation among the
observable parameters of the dilepton decay angular distribution.
The determination of invariant quantities facilitates the comparison
between measurements, and with theory, reducing the kinematic
dependence of the results to their intrinsic (and physically 
relevant) component.  Invariant relations can also be used to 
perform self-consistency checks which can expose unaccounted
systematic effects or eventual biases in the experimental analyses, a
crucial advantage given the challenging nature of quarkonium
polarization measurements due, in particular, to the difficult
subtraction of the spurious kinematic correlations induced by the
detector acceptance.

We begin with preliminary considerations on the kinematics of the
dilepton decay of inclusively produced vector mesons.  The most
general observable distribution is
\begin{align}
\begin{split}
  W(\cos \vartheta, \varphi)   \,
  & \propto \, \frac{1}{(3 + \lambda_{\vartheta})} \,
  (1 + \lambda_{\vartheta} \cos^2 \vartheta \\
  & + \lambda_{\varphi} \sin^2 \vartheta \cos 2 \varphi + \lambda_{\vartheta \varphi} \sin 2 \vartheta \cos \varphi ) \, ,
 \label{eq:ang_distr_general}
\end{split}
\end{align}
where $\vartheta$ and $\varphi$ are the
(polar and azimuthal)
angles formed by the positive lepton with, respectively, the
polarization axis $z$ and the production plane $xz$ (containing the
directions of the colliding particles and of the decaying meson).
All experimentally definable polarization axes belong to the
production plane.
A transformation from one observation frame (A) to another (B) is a
rotation around the $y$ axis by a $\delta$ angle, the parameters
changing as:
\begin{equation}
\begin{split}
\lambda^{({\rm B})}_\vartheta & =  \frac{{\lambda^{({\rm A})}_\vartheta
 - 3\Lambda }}{{1 + \Lambda }}\, ,  \quad
\lambda^{({\rm B})}_\varphi  =  \frac{{\lambda^{({\rm A})}_\varphi
 + \Lambda }}{{1 + \Lambda }}\, ,  \\
\lambda^{({\rm B})}_{\vartheta \varphi} & =  \frac{{\lambda^{({\rm
A})}_{\vartheta \varphi} \cos 2\delta  - \frac{1}{2}\, (\lambda^{({\rm
A})}_\vartheta - \lambda^{({\rm A})}_\varphi  )
\sin 2\delta }} {{1 + \Lambda }}\, , \\
\mathrm{with} \quad \Lambda & = \frac{1}{2}\, (\lambda^{({\rm A})}_\vartheta -
\lambda^{({\rm A})}_\varphi)\sin^2 \delta - \frac{1}{2}\, \lambda^{({\rm
A})}_{\vartheta \varphi} \sin 2\delta  \, . \label{eq:lambda_transf}
\end{split}
\end{equation}
Since the magnitude of the ``polar anisotropy'', $\lambda_\vartheta$,
never exceeds $1$ in any frame, we deduce the frame-independent
inequalities
\begin{equation}
|\lambda_\varphi| \le \frac{1}{2}\, (1 + \lambda_\vartheta ) \, , \quad
|\lambda_{\vartheta \varphi}| \le \frac{1}{2}\, (1 - \lambda_\varphi ) \, ,
\label{eq:triangles}
\end{equation}
which imply the bounds $|\lambda_\varphi| \le 1 $ and
$|\lambda_{\vartheta \varphi}| \le 1$.  More interestingly, we can see
that $|\lambda_\varphi| \le 0.5$ when $\lambda_{\vartheta} = 0$ and
must vanish when $\lambda_{\vartheta} \to -1$.

In general, the transformation of the polarization parameters
explicitly depends on the production kinematics.  Considering, for
example, the Collins-Soper (CS) and helicity (HX) frames, the angular
terms in Eq.~\ref{eq:lambda_transf} are
\begin{align}
\begin{split}
\sin^2\delta_{\mathrm{HX} \rightarrow \mathrm{CS}} \; & = \;
\sin^2\delta_{\mathrm{CS} \rightarrow \mathrm{HX}} \; = \;
\frac{p_\mathrm{T}^2 \, E^2}{p^2 (m^2 + p_\mathrm{T}^2)} \, , \\
\sin 2\delta_{\mathrm{HX} \rightarrow \mathrm{CS}} \; & = \; - \sin
2\delta_{\mathrm{CS} \rightarrow \mathrm{HX}} \; = \; \frac{2\, m \,
p_\mathrm{T} \, p_\mathrm{L} \, E}{p^2 (m^2 + p_\mathrm{T}^2)} \, ,
\label{eq:delta2_HX_to_CS}
\end{split}
\end{align}
where $m$, $E$, $p$, $p_\mathrm{T}$ and $p_\mathrm{L}$ are,
respectively, the mass, the energy and the total, transverse and
longitudinal momenta of the meson in the center-of-mass of the
collision. As a result, the observed quarkonium polarization has, in
general, an ``extrinsic'', frame-related, kinematic dependence,
superimposed on the ``intrinsic'' physical dependence due to the
properties of the production processes and their varying mixture.

\begin{figure}[t!]
\centering
\includegraphics[bb=0 0 556 399, width=0.38\linewidth]{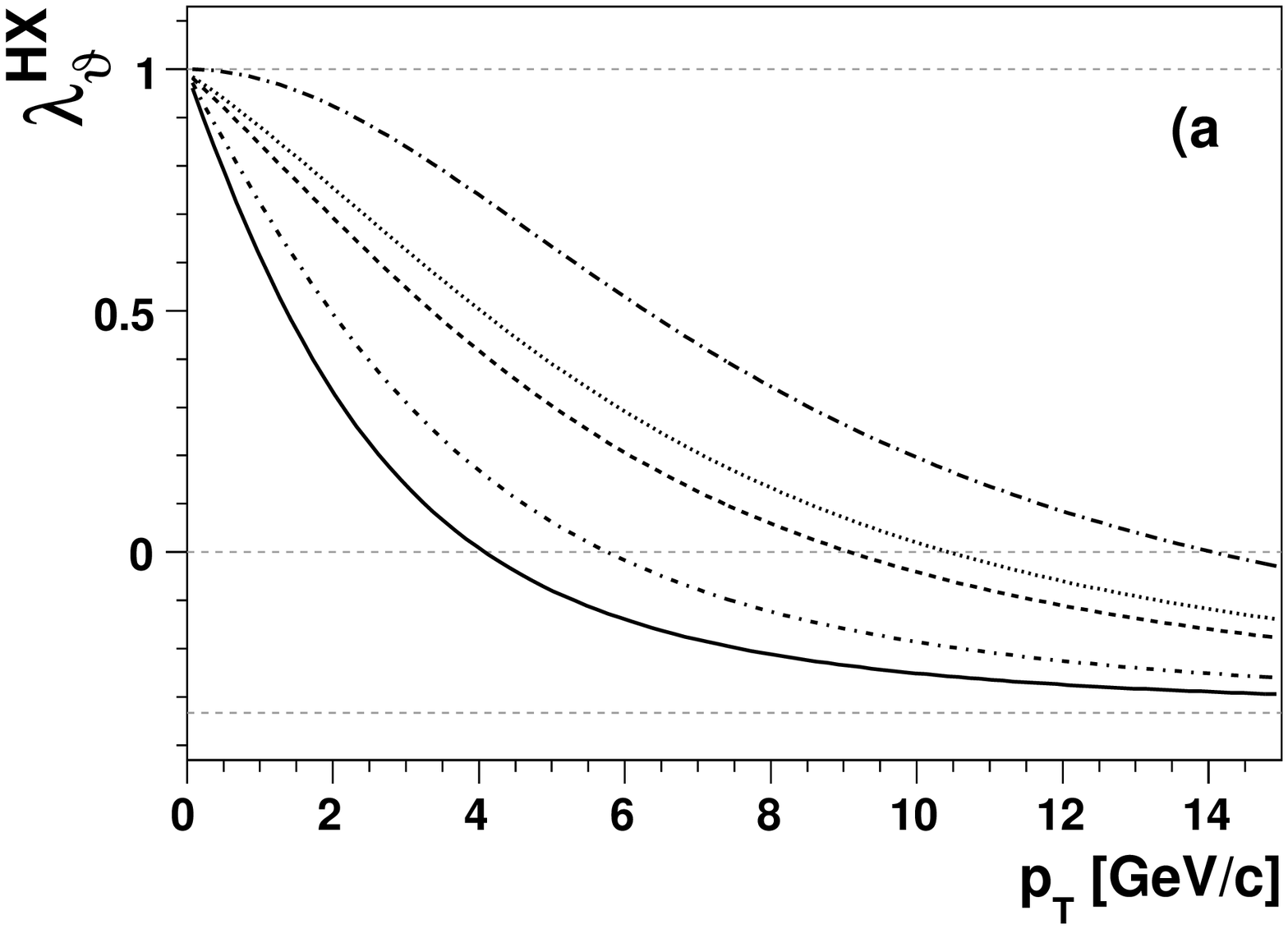}
\includegraphics[bb=0 0 556 399, width=0.38\linewidth]{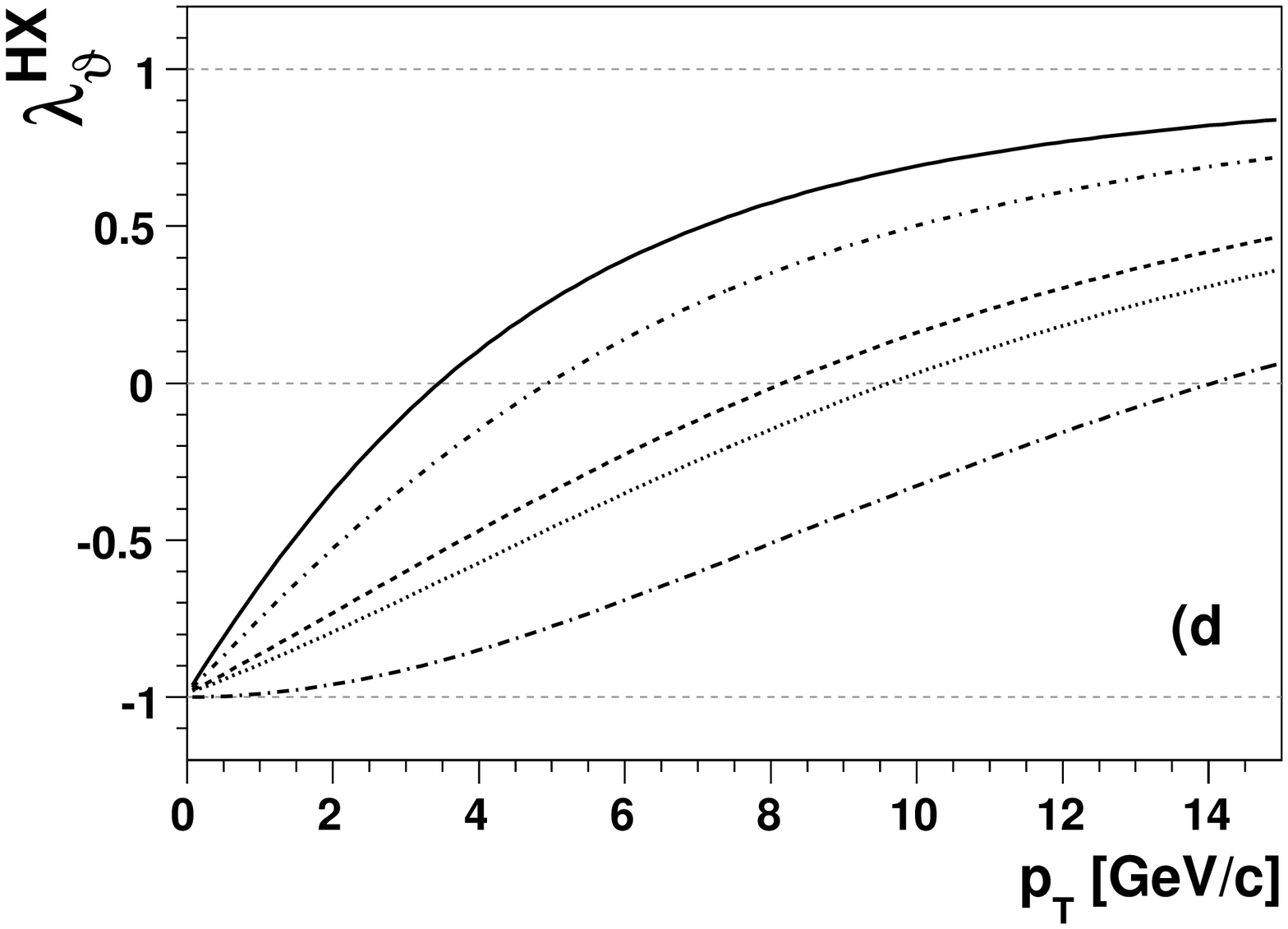}
\includegraphics[bb=0 0 556 399, width=0.38\linewidth]{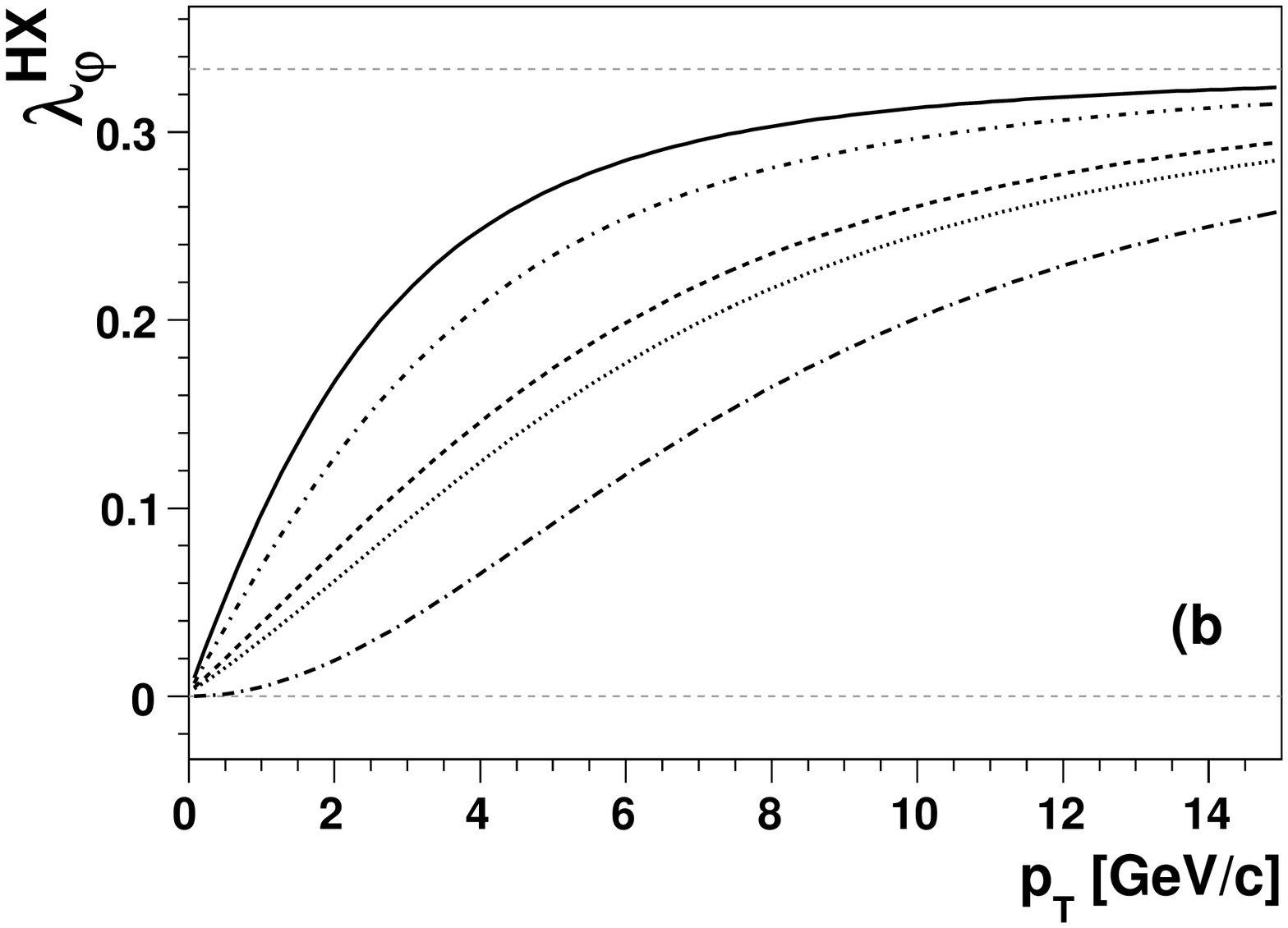}
\includegraphics[bb=0 0 556 399, width=0.38\linewidth]{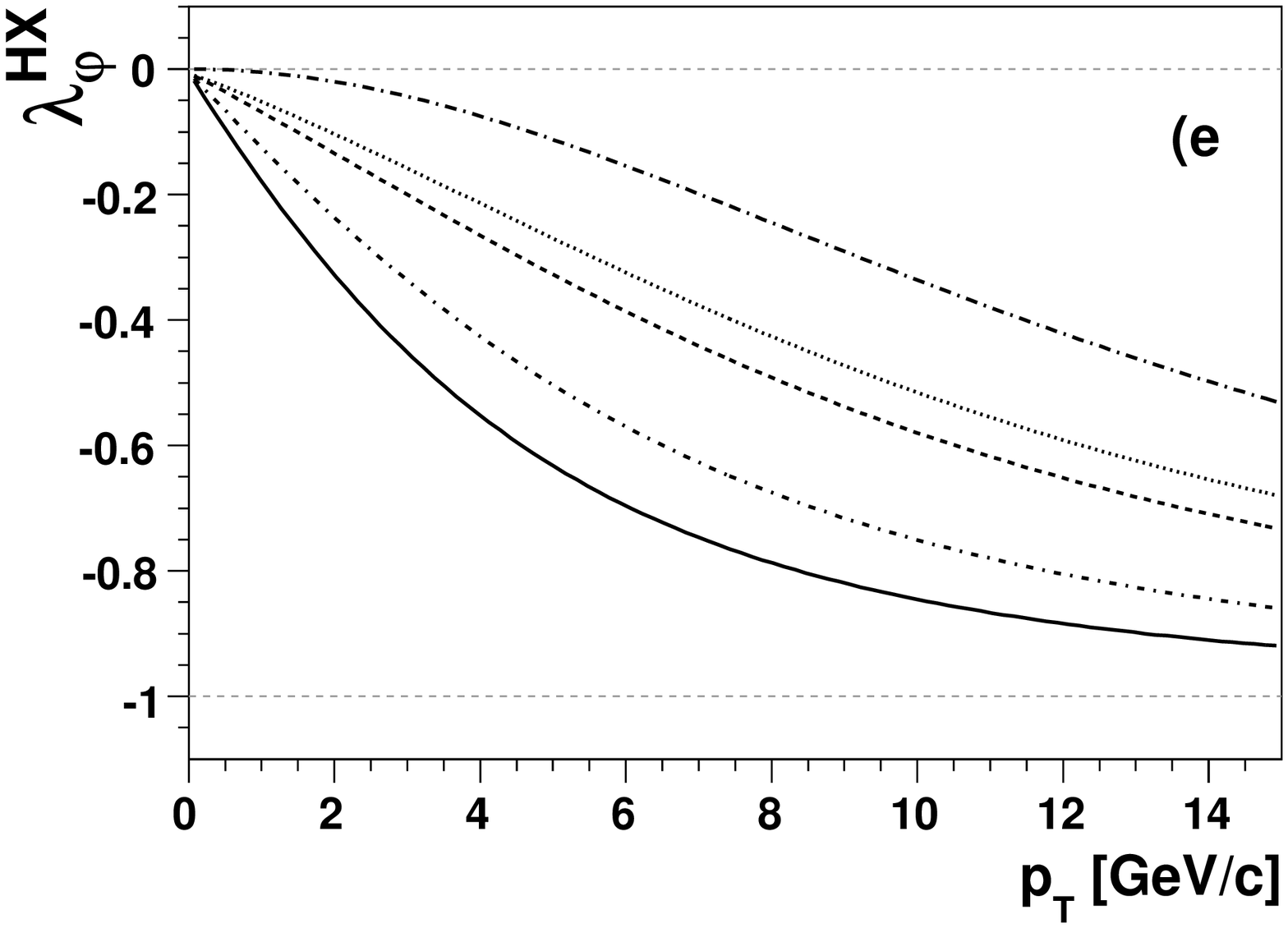}
\includegraphics[bb=0 0 556 399, width=0.38\linewidth]{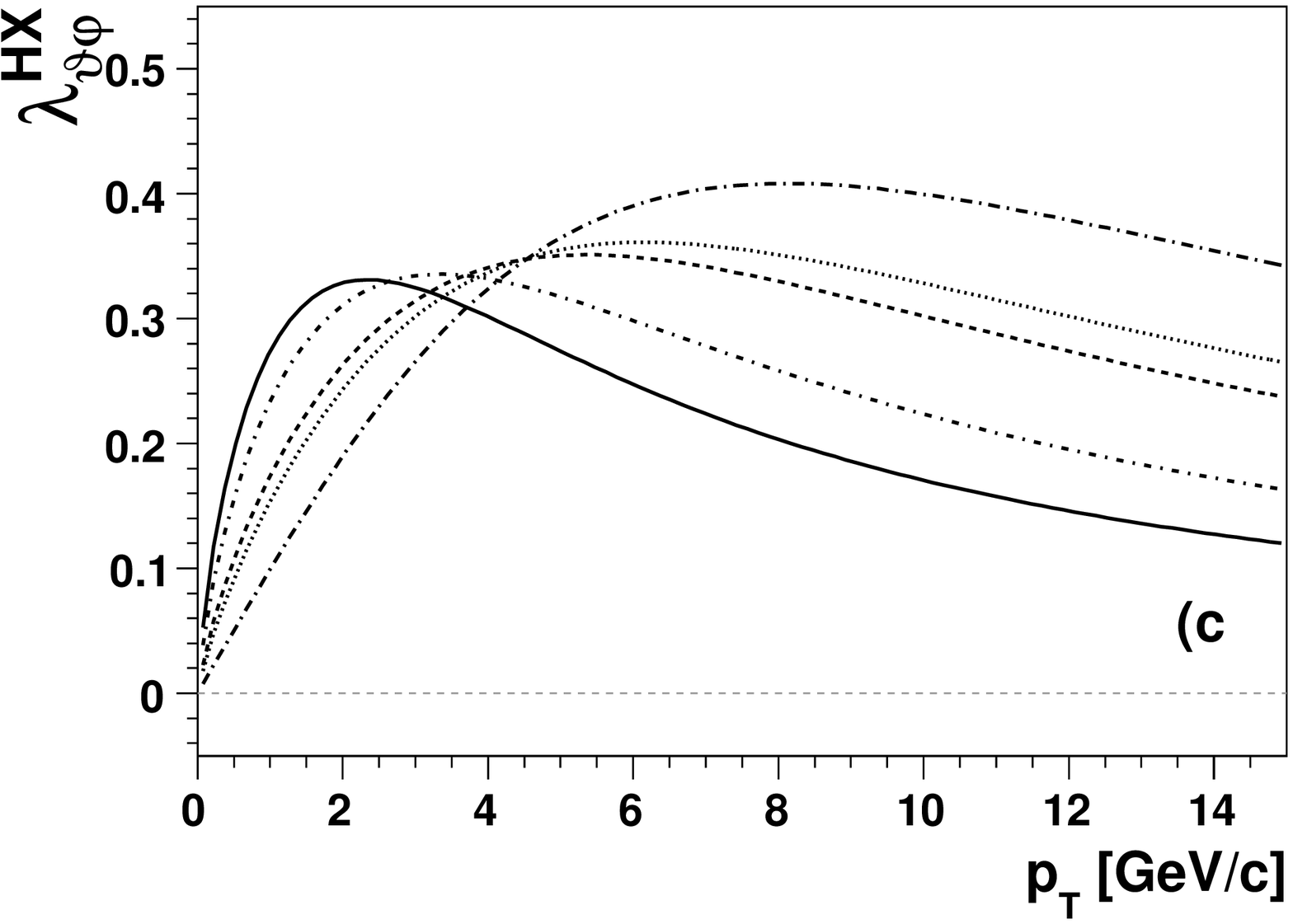}
\includegraphics[bb=0 0 556 399, width=0.38\linewidth]{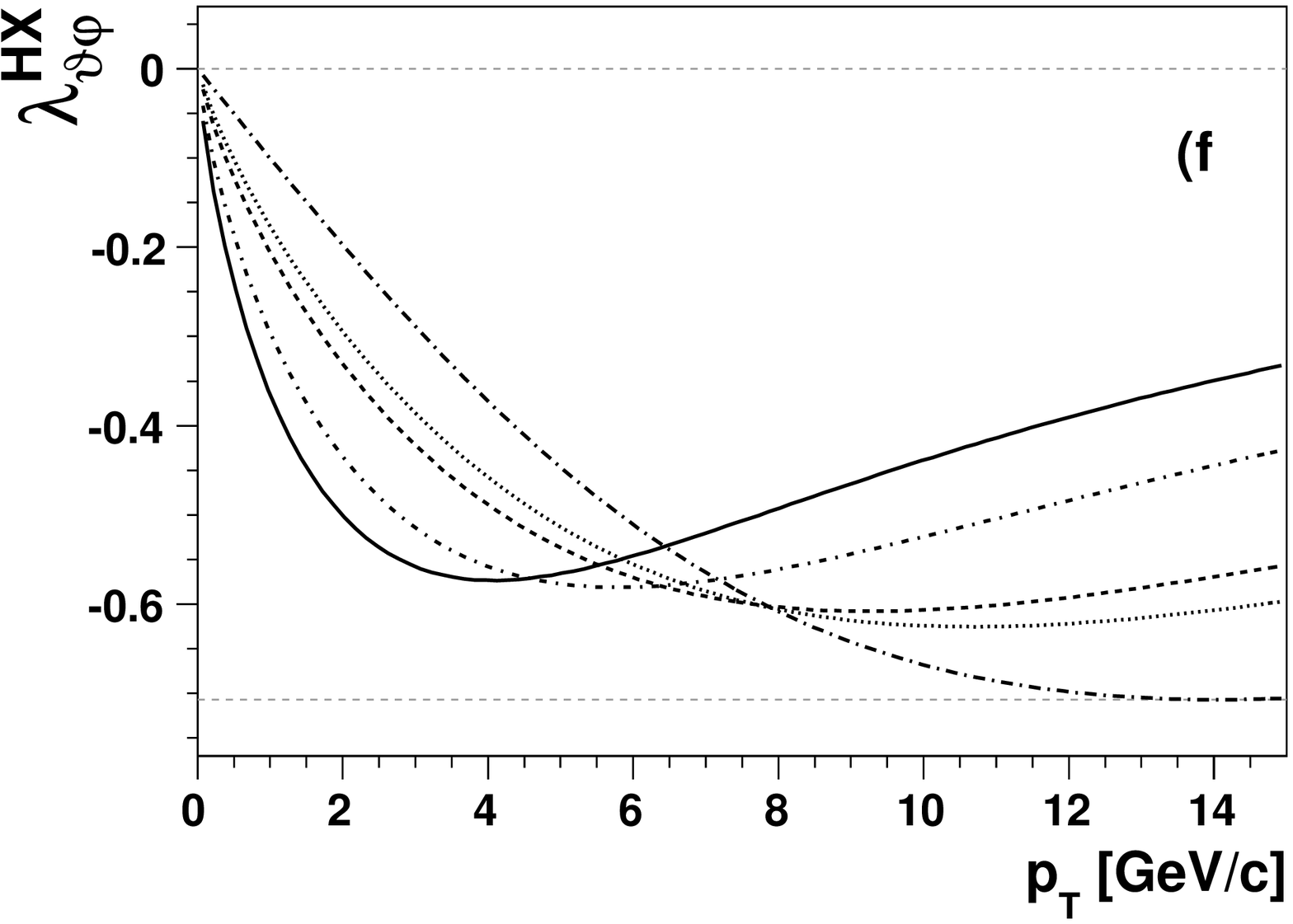}
\vglue-1mm
\caption{Kinematic dependence of the
\upsOneS\ decay angular distribution seen in the HX frame, for natural
polarizations $\lambda_{\vartheta}$\,$=$\,$+1$ (a-b-c) and
$\lambda_{\vartheta}$\,$=$\,$-1$ (d-e-f) in the CS frame. The curves correspond
to different rapidity intervals; from the
solid line: $|y| < 0.6$ (CDF), $|y| < 0.9$ (ALICE), $|y| <
1.8$ (D0), $|y| < 2.5$ (ATLAS and CMS), $2< |y| < 5$ (LHCb). For simplicity
the event populations were generated flat in rapidity. The sign of
$\lambda_{\vartheta \varphi}$ depends on the definition of the $y$ axis of the
polarization frame, here taken as $\mathrm{sign}(p_\mathrm{L}) (
\vec{P^\prime}_1 \times \vec{P^\prime}_2 ) /|\vec{P^\prime}_1 \times
\vec{P^\prime}_2|$, where $\vec{P^\prime}_{1,2}$ are the
momenta of the colliding particles in the meson's rest frame.}
\label{fig:kindep_lambda}
\end{figure}

\begin{figure}[t!]
\centering
\includegraphics[bb=0 0 556 399, width=0.38\linewidth]{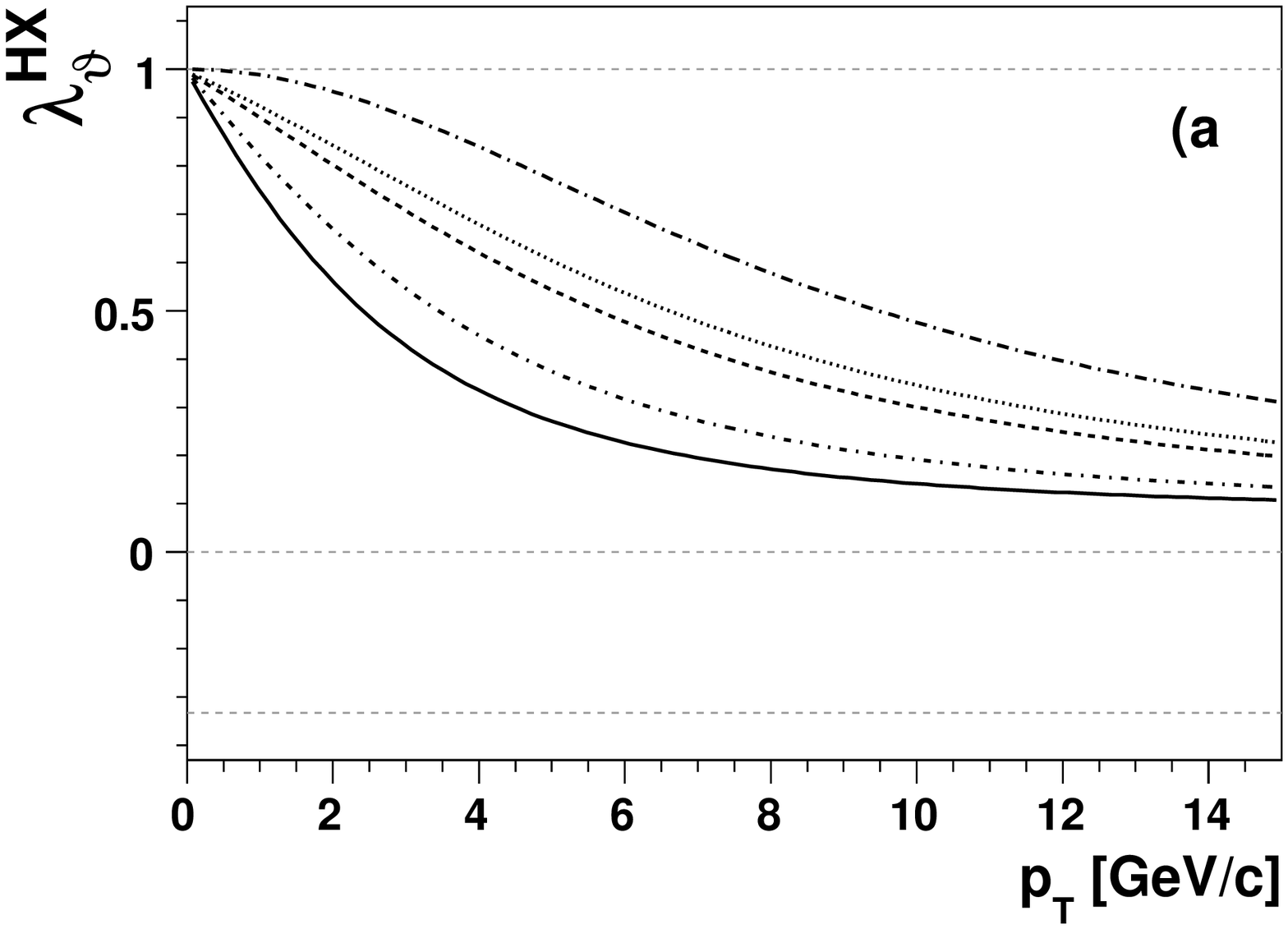}
\includegraphics[bb=0 0 556 399, width=0.38\linewidth]{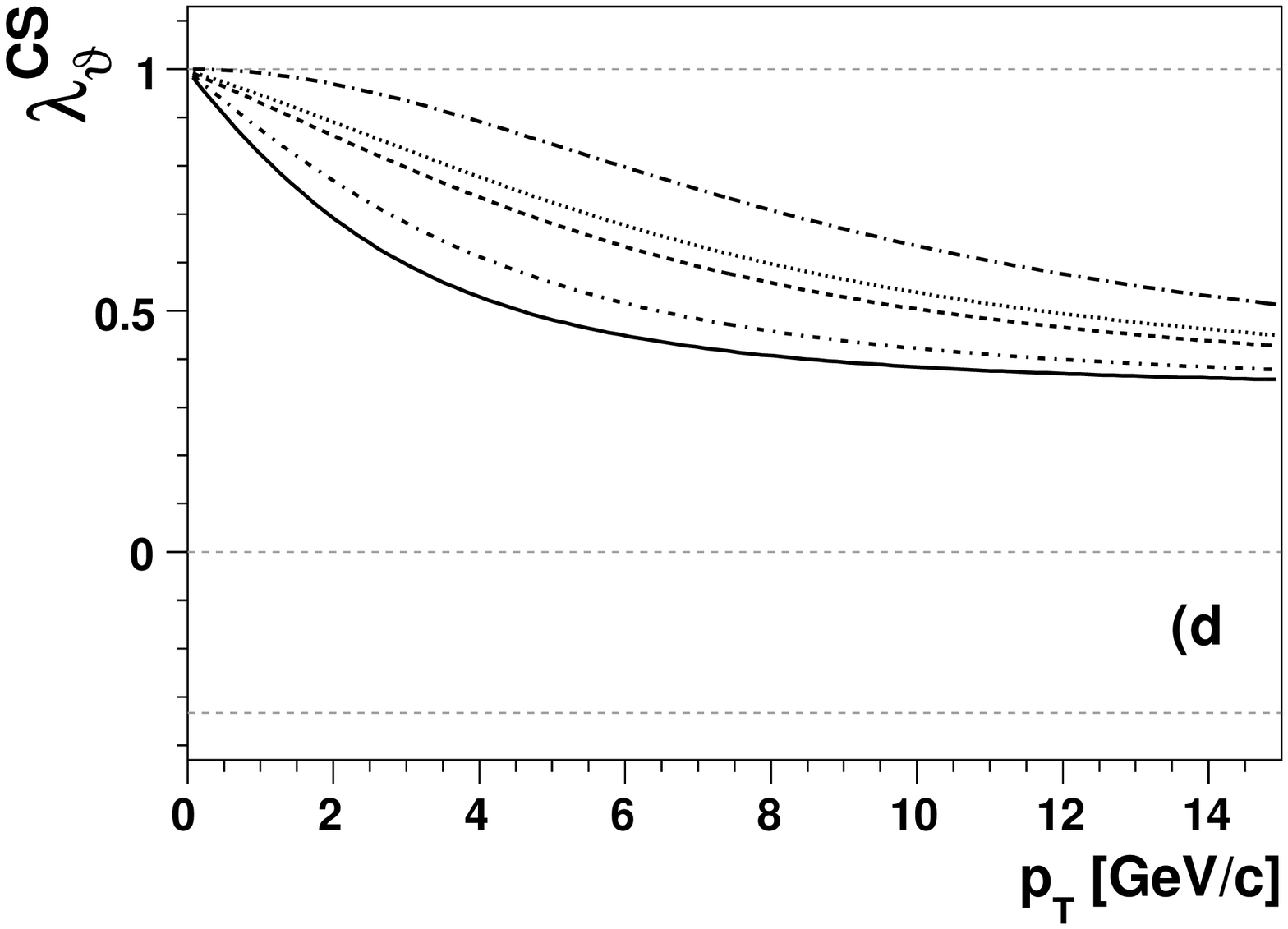}
\includegraphics[bb=0 0 556 399, width=0.38\linewidth]{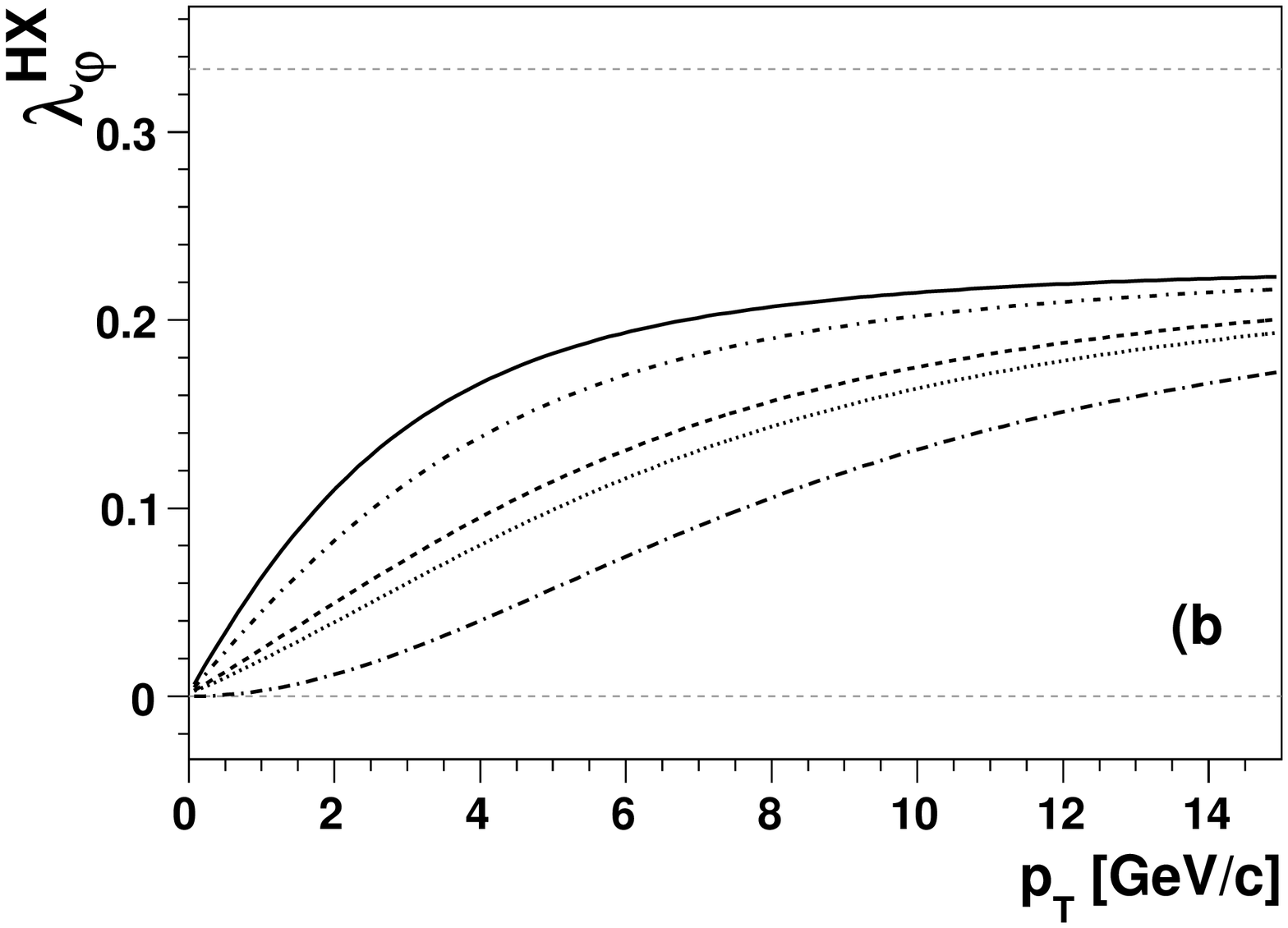}
\includegraphics[bb=0 0 556 399, width=0.38\linewidth]{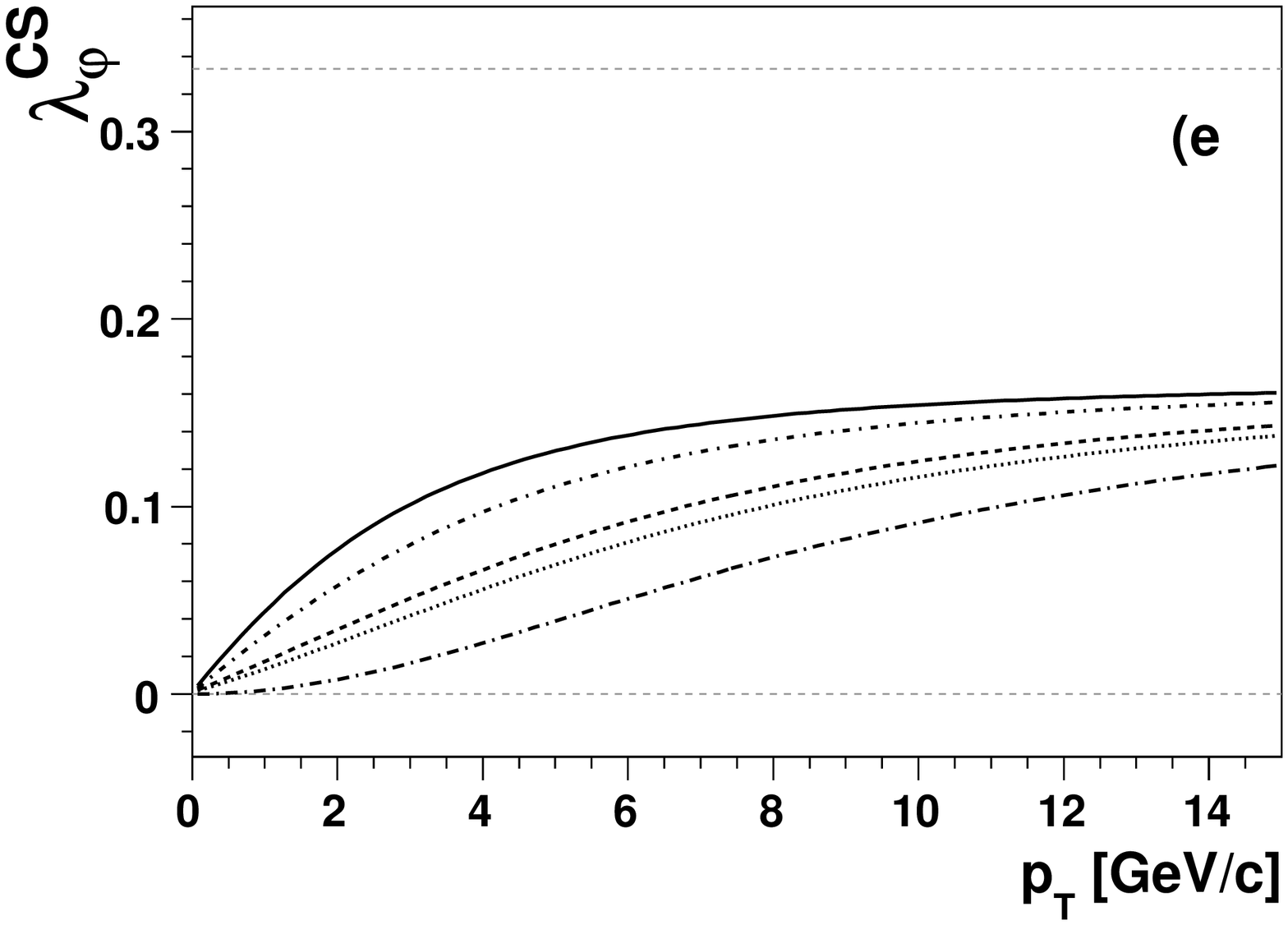}
\includegraphics[bb=0 0 556 399, width=0.38\linewidth]{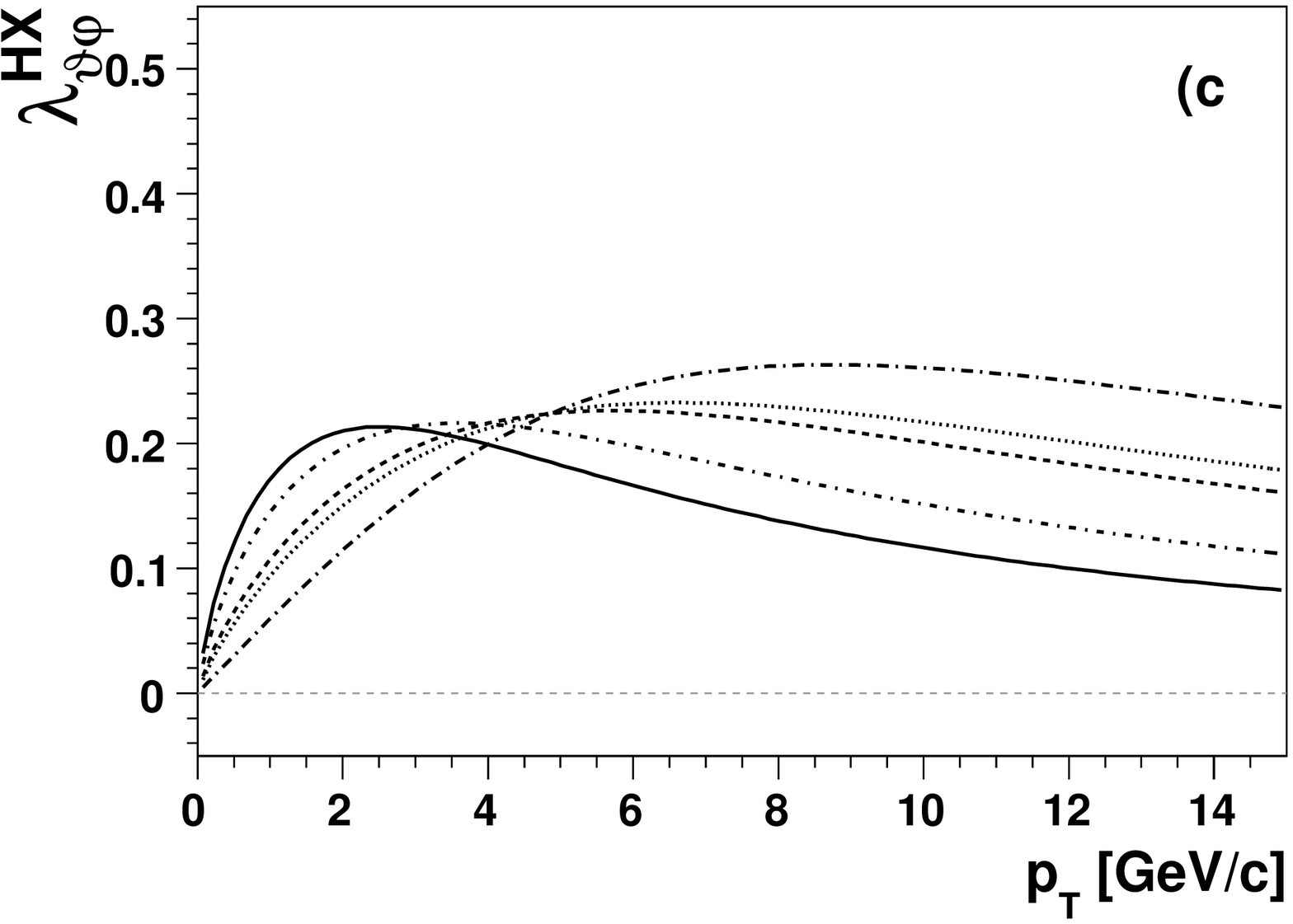}
\includegraphics[bb=0 0 556 399, width=0.38\linewidth]{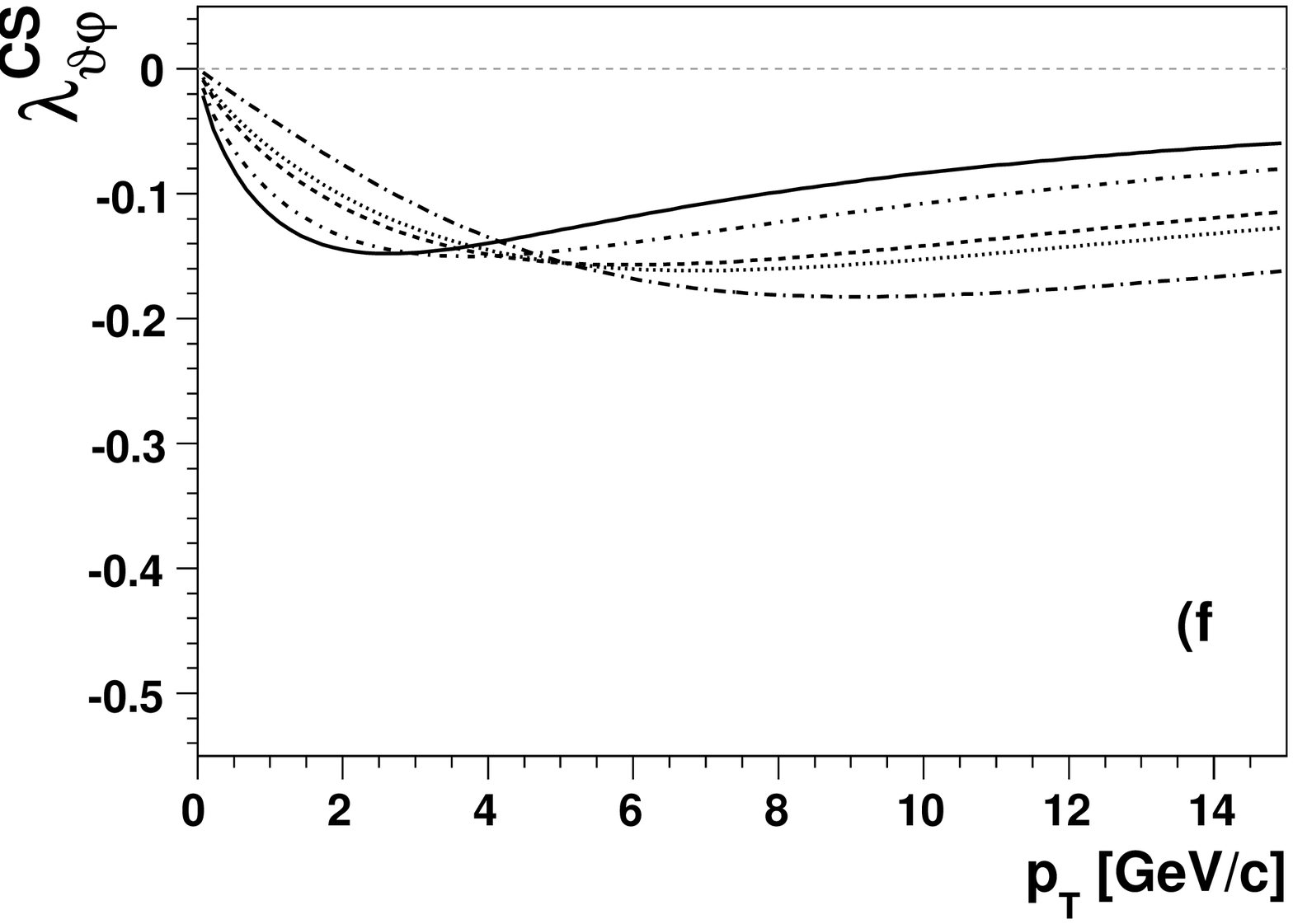}
\vglue-1mm
\caption{Kinematic dependence of the
\upsOneS\ decay angular distribution seen in the HX (a-b-c) and CS (d-e-f)
frames, when $60\%$ ($40\%$) of the events have full transverse polarization in
the CS (HX) frame. The curves represent measurements in different acceptance
ranges, as detailed in Fig.~\ref{fig:kindep_lambda}.}
\label{fig:kindep_lambda_mix}
\end{figure}

Such an extrinsic dependence can introduce artificial differences
between the polarization results obtained by experiments probing
different acceptance windows.
Figure~\ref{fig:kindep_lambda} shows how natural \upsAll\
polarizations $\lambda_\vartheta = +1$ and $-1$ in the CS frame (with
$\lambda_{\varphi} = \lambda_{\vartheta \varphi} = 0$ and no intrinsic
kinematic dependence) translate into different \pt-dependent
polarizations measured in the HX frame in different rapidity
acceptances.
In this simple case, a common choice of the ``natural'' frame (CS) by
all experiments would avoid such a misleading differentiation of
results.  In general, however, it may be impossible to find one frame
providing a simple representation of the quarkonium polarization
scenario. This is shown in Fig.~\ref{fig:kindep_lambda_mix}, where we
consider, for illustration, that $60\%$ of the \upsAll\ events have
natural polarization $\lambda_\vartheta = +1$ in the CS frame and the
remaining fraction has $\lambda_\vartheta = +1$ in the HX frame.
While the polarizations of the two event subsamples are intrinsically
independent of the production kinematics, in neither frame will
measurements performed in different transverse and longitudinal
momenta windows find identical results.

Also the comparison between experimental data and theory must consider
the experimental acceptance and efficiency.  Experiments measure the
net polarization of the specific cocktail of quarkonium events
accepted by the detector, trigger and analysis cuts.  If the
polarization depends on the kinematics, the measured angular
parameters depend on the effective population of collected events in
the probed phase space window. Two experiments covering the same
kinematic interval may find different average polarizations if they
have different acceptance shapes in that range. The problem can be
solved by presenting the results in narrow intervals of the probed
phase space. Similarly, theoretical calculations of the average
polarization in a certain experiment should consider how the momentum
distribution is distorted by its acceptance.  Alternatively, the
prediction should avoid as much as possible kinematic integrations or
provide event-level information to be embedded into Monte Carlo
simulations of the experiments.

The general frame-transformation relations in
Eq.~\ref{eq:lambda_transf} imply the existence of an invariant
quantity, definable in terms of $\lambda_{\vartheta}$,
$\lambda_{\varphi}$ and $\lambda_{\vartheta \varphi}$, in one of the
following equivalent forms:
\begin{equation}
  \mathcal{F}_{\{c_i\}} \, = \,  \frac{(3 + \lambda_\vartheta) + 
    c_1 (1 - \lambda_\varphi)}{c_2 (3 + \lambda_\vartheta) + 
    c_3 (1 - \lambda_\varphi)} \, .
\label{eq:invariants}
\end{equation}
An account of the fundamental meaning of the frame-invariance of these
quantities can be found in Ref.~\cite{bib:OtherPaper}. We will
consider here, specifically, the form
\begin{equation}
  \tilde{\lambda} \, \equiv \,  \mathcal{F}_{\{-3,0,1\}} \,
  = \, \frac{\lambda_\vartheta + 3 \lambda_\varphi }{1 -
    \lambda_\varphi} \, . \label{eq:lambda_tilde}
\end{equation}
In the special case when the observed distribution is the
superposition of $n$ ``elementary'' distributions of the kind $1 +
\lambda_\vartheta^{(i)} \cos^2\vartheta$, with event weights
$f^{(i)}$, with respect to $n$ different polarization axes,
$\tilde{\lambda}$ represents a weighted average of the $n$
polarizations, made \emph{irrespectively} of the orientations of the
corresponding axes:
\begin{equation}
 \tilde{\lambda} \; = \;
\sum_{i = 1}^{n} \frac{f^{(i)}}{3 + \lambda_{\vartheta}^{(i)}}
\, \lambda^{(i)}_{\vartheta}  
\, \bigg/ \, \sum_{i = 1}^{n} \frac{f^{(i)}}{3 +
  \lambda_{\vartheta}^{(i)}} \, .
\label{eq:lambda_tilde_meaning}
\end{equation}
The determination of an invariant quantity is immune to ``extrinsic''
kinematic dependencies induced by the observation perspective and is,
therefore, less acceptance-dependent than the anisotropy parameters
$\lambda_{\vartheta}$, $\lambda_{\varphi}$ and $\lambda_{\vartheta
  \varphi}$.
For instance, in the case illustrated in
Fig.~\ref{fig:kindep_lambda_mix}, as well as in the simpler case of
Fig.~\ref{fig:kindep_lambda}(a-b-c), any arbitrary choice of the
experimental observation frame will always yield the value
$\tilde{\lambda} = 1$, independently of kinematics. 
This particular case, where all contributing processes are
transversely polarized, is formally equivalent to the Lam-Tung
relation~\cite{bib:LamTung}, as discussed in
Ref.~\cite{bib:OtherPaper}.  Analogously, the example represented in
Fig.~\ref{fig:kindep_lambda}(d-e-f), or any other case where all
polarizations are longitudinal, yields $\tilde{\lambda} = -1$.

The existence of frame-invariant parameters also provides a useful
instrument for experimental analyses. Checking, for example, that the
same value of an invariant quantity (Eq.~\ref{eq:invariants}) is
obtained (within systematic uncertainties) in two distinct
polarization frames is a non-trivial verification of the absence of
unaccounted systematic effects.
In fact, detector geometry and/or data selection constraints strongly
polarize the reconstructed dilepton events.  Background processes also
affect the measured polarization, if not well subtracted.
The spurious anisotropies induced by detector effects and background
do not obey the frame transformation rules characteristic of a
physical $J=1$ state.
If not well corrected and subtracted, these effects will distort the
shape of the measured decay distribution differently in different
polarization frames.  In particular, they will violate the
frame-independent relations between the angular parameters. 
Any two physical polarization axes (defined in the rest frame of the
meson and belonging to the production plane) may be chosen to perform
these tests. The HX and CS frames are ideal choices at high \pt, where
they tend to be orthogonal to each other (in
Eq.~\ref{eq:delta2_HX_to_CS}, $\sin^2 \delta \rightarrow 1$ for
$p_\mathrm{T} \gg m$). 
At low \pt, where the difference between the two frames vanishes, any
of the two and its exact orthogonal may be used to maximize the
significance of the test.
Given that $\tilde{\lambda}$ is ``homogeneous'' to the anisotropy
parameters, the difference $\tilde{\lambda}^{({\rm B})} -
\tilde{\lambda}^{({\rm A})}$ between the results obtained in two
frames provides a direct evaluation of the level of systematic errors
not accounted in the analysis.

We conclude with a summary of our messages.  Choosing a given
polarization axis in an experimental analysis or theoretical
calculation has a radical effect on the magnitudes and signs of the
polar and azimuthal anisotropies observed in the dilepton decay
distribution: all terms of the distribution must be determined to
provide unambiguously interpretable physical information.
Rotational invariance imposes frame-invariant constraints on the polar
and azimuthal anisotropy parameters and, for any mixture of production
mechanisms in a given kinematic condition, there exists an invariant
relation depending on one frame-independent parameter.
Measurements of the anisotropy parameters are necessarily affected by
the fact that the transformation from one frame to another is an
explicit function of the production kinematics. These effects may
result in a misleading interpretation of the measurements.
Reporting polarization results in terms of invariant quantities
facilitates the comparison between different measurements, and with
theory, reducing the kinematic dependencies to their intrinsic (and
physically most relevant) component. The invariant relation can also
be used in the data analyses to perform self-consistency checks which
can expose unaccounted detector effects or eventual biases.

\bigskip

P.F.\ and J.S.\ acknowledge support from 
Funda\c{c}\~ao para a Ci\^encia e a Tecnologia,
%
Portugal, under contracts SFRH/BPD/42343/2007 and CERN/FP/109343/2009.


\end{document}